\documentclass[fleqn,usenatbib,usedcolumn]{mnras}
  \usepackage{amsmath}
  \usepackage{bm}
  \usepackage{upgreek}
   \usepackage{hyperref}
   
   \usepackage{graphicx,color,amsmath,amssymb,amsfonts}
   
   \usepackage[british]{babel}             
   \usepackage{txfonts}                    
   \let\la=\lesssim 
   \usepackage{graphicx}                   

   \hypersetup{pdfauthor={Jos\'e Fonseca, Roy Maartens, M\'ario G. Santos},
               pdftitle={Synergies between intensity maps of Hydrogen lines},
               pdfkeywords={cosmology: large-scale structure of the Universe, cosmological parameters, miscellaneous},
               bookmarksnumbered=true}
   \hypersetup{colorlinks=true,
            linkcolor=blue,
            citecolor=blue,
            filecolor=blue,
            urlcolor=blue}

\usepackage{xcolor}
 
\newcommand{\be}{\begin{eqnarray}}
\newcommand{\ee}{\end{eqnarray}}
\newcommand{\bea}{\begin{eqnarray}}
\newcommand{\eea}{\end{eqnarray}}
\newcommand{\bfig}{\begin{figure}}
\newcommand{\efig}{\end{figure}}
\newcommand{\nn}{\nonumber}

\newcommand{\ylm}[1]{Y_{\ell m}(\boldsymbol #1)}

\def\p{\partial}
\def\fsky{f_{\rm sky}}
\def\z{\l z\r}
\def\odm{\Omega_{\rm cdm}}
\def\ob{\Omega_{\rm b}}
\def\cs{CAMB$\_$sources}
\def\fnl{{f_{\rm NL}}}
\def\l{\left(}
\def\r{\right)}
\def\s{{\sigma}}
\def\D{\Delta}
\def\a{{\alpha}}
\def\d{\mathrm{d}}
\def\H{{\mathcal H}}
\def\alm{a_{\ell m}}
\def\la{\lambda}

\title[Multi-tracing hydrogen lines]{Synergies between intensity maps of hydrogen lines}
\author[Fonseca et al.]{Jos\'e Fonseca$^1$\thanks{josecarlos.s.fonseca@gmail.com}, Roy Maartens$^{1,2}$,  M\'ario G. Santos$^{1,3}$ \\
$^1$ Department of Physics and Astronomy, University of the Western Cape, Cape Town 7535, South Africa\\
$^2$ Institute of Cosmology \& Gravitation, University of Portsmouth, Portsmouth PO1 3FX, UK\\ 
$^3$ SKA SA, The Park, Park Road, Cape Town 7405, South Africa\\
}
\date{}

\pagerange{\pageref{firstpage}--\pageref{lastpage}}

\begin{document}

\label{firstpage}

\maketitle

\begin{abstract}
We study synergies between \textsc{Hi} 21cm and H$\alpha$ intensity map observations, focusing on SKA1-like and SPHEREx-like surveys. We forecast how well such a combination can measure features in the angular power spectrum on the largest scales, that arise from primordial non-Gaussianity and from general relativistic effects. For the first time we consider Doppler, Sachs-Wolfe and integrated SW effects separately. We confirm that the single-tracer surveys on their own  cannot detect general relativistic effects and can  constrain the non-Gaussianity parameter $\fnl$ only slightly better than Planck. Using the multi-tracer technique, constraints on $\fnl$ can be pushed down to $\sim1$. Amongst the general relativistic effects, the Doppler term is detectable with the multi-tracer. The Sachs-Wolfe terms and the integrated SW effect are still not detectable.

\end{abstract}





\section{Introduction}

There has been a growing interest in the field of line intensity mapping (IM) for cosmological measurements, such as the 3-dimensional (3D) large-scale structure of the Universe across cosmic time. Instead of resolving each individual galaxy, IM integrates over all emission inside the {voxel} (3D pixel). Hence, fluctuations in observed intensity come from fluctuation in the number of sources inside the voxel. Since we expect to have several galaxies in each pixel, the signal-to-noise should be higher than for standard galaxy surveys with a threshold magnitude. The technique has another advantage with respect to conventional photometric and spectroscopic galaxy surveys -- one can probe larger sky areas with higher redshift resolution more speedily. The downside comes at the loss of small scale information. While galaxy surveys rely on clean galaxy number counts, i.e., either a galaxy is detected or not, IM has to deal with interloping line emission or other types of contaminants. 

A lot of focus has been applied to the \textsc{Hi} 21cm line as a tracer of the dark matter distribution in the late Universe \citep{Battye:2004re,  Chang:2007xk,2008PhRvL.100p1301L,Bharadwaj:2008yn,Bagla:2009jy, Seo:2009fq,Chang:2010jp,2012A&A...540A.129A,2013MNRAS.434L..46S,2013MNRAS.434.1239B, Bull:2014rha}. In fact, some observations have already been made \citep{2011AN....332..637K}, with a detection in cross-correlation \citep{Chang:2010jp}. A large-area \textsc{Hi} intensity mapping survey has been proposed for SKA1-MID \citep{Santos:2015bsa} and for its precursor MeerKAT \citep{Santos:2017aa}. An advantage of using the \textsc{Hi} line is that it has very little background and foreground contamination from other lines, but there is major synchrotron contamination from our Galaxy. 

Some authors have considered other lines as cosmological probes at low redshift, such as \textsc{Ly$\a$} \citep{Pullen:2013dir}, \textsc{Cii} and other far infra-red emission lines \citep{Uzgil:2014pga} and the \textsc{CO} rotation emission lines \citep{Breysse:2014uia}. \citet{Fonseca:2016qqw} performed a systematic study of all the lines (besides \textsc{Hi}) that can in principle be used for IM, including \textsc{H$\a$} and fainter optical and far infra-red lines, to assess which had the best prospects for use as IM probes and to forecast which lines can provide measurements of the 3D power spectrum in the future. A more comprehensive summary of the current status of line IM is given by \citet{Kovetz:2017agg}.

Planned galaxy surveys will probe ever larger volumes of the Universe. This will deliver not only tighter and tighter constraints on the standard cosmological model, but also allow us to go further and test other features, such as primordial non-Gaussianity (PNG, via the parameter $\fnl$), which induces a scale-dependent correction to the bias of any dark matter tracer \citep{Dalal:2007cu,Matarrese:2008nc}. Current Planck bounds on $\fnl$ are the state of the art but do not provide a clear measurement: $\fnl\simeq0.8\pm 5.0$ \citep{Ade:2015ava}. Future surveys should deliver improvements on CMB constraints by using the scale-dependent bias of tracers \citep[see, e.g.][]{Giannantonio:2011ya,Camera:2013kpa,Camera:2014bwa,Alonso:2015uua,Raccanelli:2015vla}. The first target is $\sigma(\fnl)<1$, the threshold to start distinguishing between single-field and multi-field inflation \citep[see, e.g.][]{dePutter:2016trg}. However, the forecasts show that single-tracer surveys cannot reach this level of precision, even with a perfect experiment, because of cosmic variance. 

On the ultra-large scales ($k<k_{\rm eq}$) where the PNG signal in IM is strongest, general relativistic (GR) effects on the observed brightness temperature \citep{Hall:2012wd} are also strongest. Neglecting these effects would bias measurements of $\fnl$ in future surveys \citep{Camera:2014sba}.  Furthermore, it would discard additional information and potential tests on such uncharted scales. 
In IM, the GR effects include Doppler, Sachs-Wolfe (SW) and integrated Sachs-Wolfe (ISW) terms, whereas number counts have in addition lensing\footnote{Note that the GR lensing contribution to number counts can be signifcant on sub-equality scales.} and time-delay terms \citep{Yoo:2010ni,Challinor:2011bk,Bonvin:2011bg}. Several papers have studied how well future galaxy and IM surveys will be able to distinguish GR corrections from the dominant contributions to the density contrast \citep[see e.g.][]{Yoo:2012se,Alonso:2015uua,Camera:2015fsa,Raccanelli:2015vla}. Because of the same cosmic variance problem that limits $\sigma(\fnl)$, GR effects (other than lensing) cannot be detected in single-tracer surveys.

Cosmic variance is indeed sampling variance of cosmological scales. Since we only have one realisation of the Universe, ever larger scales have ever less repetitions in the cosmological statistical ensemble at that scale. Despite this, a way has been proposed to beat down cosmic variance when measuring effects in tracer-dependent bias-like parameters. The multi-tracer (MT) technique  to combines two or more tracers of the underlying dark matter field \citep{Seljak:2008xr,McDonald:2008sh,Hamaus:2011dq,Abramo:2013awa,Abramo:2015iga} to cancel sample variance. This technique is opening up a new window on constraining and measuring ultra-large scale effects in the 3D power spectrum with next-generation surveys of the large-scale structure \citep{Yoo:2012se,Ferramacho:2014pua, Yamauchi:2014ioa,Alonso:2015sfa,Fonseca:2015laa,Fonseca:2016xvi,Abramo:2017xnp}. In addition to reducing cosmic variance, the MT technique contains cross-correlations (between tracers and redshift bins) that help to reduce individual systematics and foreground/background residuals. 

In this paper we consider for the first time the potential synergies between two IM surveys using the MT technique: a \textsc{Hi}  survey like the survey  planned for Phase 1 of the SKA\footnote{www.skatelescope.org} \citep{Maartens:2015mra,Santos:2015bsa}, and a H$\alpha$ survey like the survey envisaged in the proposal for SPHEREx\footnote{http://spherex.caltech.edu}  \citep{Dore:2014cca}.  We investigate how the results depend on sky area and the assumed noise of the experiments, as well as the redshift resolution of the combined survey. Our forecasts show that in principle, the Doppler term can  be measured at $\sim3\sigma$, and $\sigma(\fnl)\sim 1$ is achievable. As we will explain, the smallness of the bias ratio of \textsc{HI} and H$\a$ undermines the constraining power of these two tracers on $\fnl$ and GR effects. 

The paper is organised as follows: in \S \ref{sec:angpow} we review the 3D angular power spectrum for IM, including all large-scale effects; we  describe the experimental specifications of the surveys  in \S \ref{sec:surveys}; \S \ref{sec:fisher_forecasts} reviews the Fisher forecast analysis and presents the chosen observational strategy; the results are presented in \S \ref{sec:results} and discussed in \S \ref{sec:discussion}.

\section{Angular power spectrum for intensity maps}\label{sec:angpow}

Let $\Delta^{W_A}(z_i,\bm{n})$ be the observed contrast of intensity or temperature fluctuations for tracer A in  direction $\boldsymbol n$ and in a  redshift bin (with window $W_A$) centred on $z_i$. We decompose each sky map in a spherical harmonic basis $ \ylm{n}$ and use the amplitudes $\alm^{W_A} (z_i)$ as our observational estimator, where 
\be
\Delta^{W_A}(z_i, \boldsymbol n)= \sum^{\infty}_{\ell = 0} \sum^{\ell}_{m = -\ell} \alm^{W_A}(z_i) \ylm{n}\,.
\ee
Since the universe is statistically homogeneous and isotropic, $\langle a_{\ell m} \rangle = 0$ at every redshift. Assuming that the $\alm$ are Gaussian-distributed, the information about the large-scale structure is encoded in the covariance of the $\alm$, i.e., the angular power spectra $C^{AB}_\ell$:
\be
\Big\langle a^{W_A}_{\ell m}(z_1)\ a^{*\,W_B}_{\ell' m'}(z_2)\Big\rangle = \delta_{\ell \ell'}\ \delta_{m m'}\ C^{AB}_\ell\big(z_1,z_2\big)\,.
\ee
We can write the angular power as  \citep{Challinor:2011bk} 
\be \label{eq:clgeneral}
C^{AB}_\ell \big( z_i,z_j\big)=4\pi\!\!\int\!\!\d \ln k\,\D_\ell^{W_A}\big( z_i,k\big) \D_\ell^{W_B}\big( z_j,k\big)\ \mathcal P (k),
\ee
where 
\bea
\mathcal P (k)=A_s\ \l \frac k{k_0}\r^{n_s-1},
\eea
is the primordial power spectrum of the curvature perturbation. The pivot scale is $k_0=0.05\,$Mpc$^{-1}$, $A_s$ is the amplitude and $n_s$ is the spectral index. The $\D^{W_A}_\ell(z_i,k)$ are the observed transfer functions at  comoving scale $k$, which include the observational effects from the window function $W_A$ and redshift distribution of sources $p^A$:
\bea \label{transf}
\D_\ell^{W_A}\big( z_i,k\big)=\int\!\! \d z\,p^A(z)\ W_A\big(z_i,z\big)\ \D_\ell^A\big(z,k\big)\,.
\eea
Here $\D_\ell^A(z,k)$ is the theoretical transfer function. If we were able to measure infinitesimal shells in redshift then our window function would just be a Dirac delta function and  the observed and theoretical transfer functions would be the same. The product $p^A W_A$ is in effect a probability distribution function for a source to contribute to the signal in the bin, and is thus normalised to unity: $\int\!\d z\,p^A(z)W_A(z_i,z)=1$ for all $z_i$. 

The full expression (including all GR corrections) for the theoretical transfer function of a map of intensity is \citep{Hall:2012wd}
\bea
\D^{A}_\ell &=& \D_\ell^{A\,{\rm  standard}}+\D_\ell^{A\,{\rm Doppler}}+ \D_\ell^{A\,{\rm SW}}+ \D_\ell^{A\,{\rm ISW}}\,, \nn \\
\D_\ell^{A\,{\rm standard}}&=& b^A\delta_k^{\rm s}\, j_\ell\l k\chi\r+\frac {k }{\H}v_k\,j_\ell''(k\chi)\,,\nn\\
\D_\ell^{A\,{\rm Doppler}}&=& \l 2-b^A_e+\frac{\dot\H}{\H^2}\r v_k\, j'_\ell(k\chi)\,, \nn\\
\D_\ell^{A\,{\rm SW}}&=&\left[\l b^A_e-3\r\frac {\H }{k}v_k+
\l 3-b^A_e+\frac{\dot\H}{\H^2}\r \psi_k+\frac{\dot\phi_k}{\H}\right]  j_\ell\l k\chi\r \,, \nn\\
\D_\ell^{A\,{\rm ISW}}&=&\l 2-b^A_e+\frac{\dot\H}{\H^2}\r \int_0^{\chi}\d\tilde\chi\l\dot\phi_k+\dot\psi_k\r j_\ell\l k\tilde\chi\r \label{eq:angIM}\,.
\eea
To simplify notation we suppressed the redshift and $k$ dependence (except in the spherical Bessel functions). Here ${\cal H}=\dot a/a$ is the conformal Hubble parameter and $\chi$ is the comoving line-of-sight distance. Equation \eqref{eq:angIM} is in Newtonian gauge, but the comoving-synchronous density contrast $\delta^{\rm s}_k$ arises
from the relativistic definition of clustering bias \citep{Challinor:2011bk,Bonvin:2011bg,Bruni:2011ta,Jeong:2011as}. The other perturbations are the peculiar velocity $v_k$ (where $ikv_k$ is the velocity divergence), and the metric perturbations, defined by 
\bea 
{\d}s^2=a^2\ \Big[-\big(1+2\psi\big)\ {\d}\eta^2 + \big(1-2\phi\big)\ {\d}\boldsymbol{x}^2\Big].
\eea
The evolution bias $b^A_e(z)$ accounts for redshift evolution of the sources: 
 \bea \label{be}
b^A_e= -\frac{\p\ln n^A}{\p\ln (1+z)}\,,
\eea
where $n^A$ is the comoving number density of A-sources.

The second line of \eqref{eq:angIM} contains the standard density + Kaiser redshift-space distortions (RSD) terms. The third line is the Doppler term. 

The fourth line is the SW contribution from the metric potentials $\phi,\psi$. We have included in this contribution the relativistic correction to clustering bias, which is the first term in the square brackets, since the velocity potential $v_k/k$ can be rewritten in terms of $\phi,\psi$ as follows.
Using the Poisson equation, $-k^2\phi_k=3\Omega_m \H^2\,\delta^{\rm s}_k/2$, and the continuity equation,
\bea \label{eq:linvel}
ikv_k=f{\H}\, \delta^s_k\,,\quad f=-\frac{\d \ln D}{\d \ln (1+z)}\,,
\eea
where $f$ is the growth rate, we find that
\bea 
\frac{\cal H}k v_k=-i \frac{2f}{3\Omega_m}\,\phi_k\,.
\eea
The $\dot{\phi}_k$ term in the SW contribution can be rewritten as
\be
\frac{\dot{\phi}_k}{\cal H}=f\phi_k-\psi_k\,,
\ee
using the momentum constraint, $\dot\phi_k+\H\psi_k=(3i\Omega_m\H^2/2)v_k/k$. 

The last line of \eqref{eq:angIM} is the ISW effect in the observed clustering. The last three lines together are the GR corrections, which are only relevant on ultra-large scales. Furthermore the ISW term is only important in the presence of dark energy, i.e., at low redshift. 

In the concordance model $\psi=\phi$ and $\dot\H/\H^2=1-3\Omega_m/2$, so that we can rewrite the GR parts as
\bea
\D_\ell^{A\,{\rm Doppler}} &=&\! \l 3-\frac32\Omega_m-b^A_e\r v_k\, j'_\ell(k\chi)\,, \label{dop2}\\
\D_\ell^{A\,{\rm SW}}&=&\!\left[\left(3-b^A_e\right)\left(1+i\frac{2f}{3\Omega_m}\right)+f-\frac32\Omega_m\right] \phi_k\, j_\ell\l k\chi\r \!, \label{sw2}\\ \label{isw2}
\D_\ell^{A\,{\rm ISW}}&=&\!\l 6-3\Omega_m-2b^A_e\r \int_0^{\chi}\d\tilde\chi\,\dot\phi_k\, j_\ell\l k\tilde\chi\r \,.
\eea

Local primordial non-Gaussianity alters the clustering bias on ultra-large scales \citep{Dalal:2007cu,Matarrese:2008nc}:
\bea\label{eq:bng}
 b^A(z,k)=b^A_G(z)+{D(z_{\rm dec})\big(1+z_{\rm dec}\big)}\, {\fnl}\, \frac{3\big[ b^A_{G}(z) -1\big]\delta_c\Omega_{m0} H_0^2 }{D(z) T(k) \, k^2},
\eea
where $b_{G}$ is the linear Gaussian bias, $\delta_c\simeq 1.69$ is the critical matter density contrast for spherical collapse, $T$ is the matter transfer function (normalised to 1 on large scales) and $D$ is the growth factor (normalised to 1 at $z=0$). The factor $D(z_{\rm dec})(1+z_{\rm dec})\approx 1.27$ ensures that we use the CMB definition of $\fnl$.

On ultra-large scales $T(k)\approx 1$, and the bias is dominated by the $\fnl$ correction with a $k^{-2}$ dependence. The same $k$-dependence holds for the SW terms in \eqref{eq:angIM}. On the other hand, the Doppler term has a $k^{-1}$ dependence from \eqref{eq:linvel}. 
We therefore expect degeneracies between $\fnl$ and the potential terms, but $\fnl$ and the Doppler term should be fairly independent.

\section{Surveys}\label{sec:surveys}

There are 3 intrinsic tracer quantities that we need to determine: the clustering bias, the evolution bias and the redshift distribution of sources. In addition, we need to specify the instrumental noise. We will neglect shot-noise of the lines, which is a safe assumption for \textsc{Hi} \citep{2011ApJ...740L..20G} and H$\alpha$ \citep{Fonseca:2016qqw}. The angular power spectrum of instrumental noise for IM experiments is similar to that of CMB experiments \citep{Knox:1995dq}. It is given by $\mathcal N_{\ell}=\sigma^2\, \Omega_{\rm pixel}$, where $\sigma$ is the sensitivity of the experiment for a chosen survey strategy and $\Omega_{\rm pixel}$ is the pixel area given by the experiment's angular resolution. Note that the noise angular power does not depend on the angular scale $\ell$. Following the approach of \citet{Knox:1995dq}, beam effects should be included in the signal, not in the noise. 
While the two approaches are equivalent for a single tracer in a single redshift bin, this does not hold for multiple tracers or multiple redshift bins. The effect of the beam is to reduce power at high $\ell$. For the purposes of this paper, we  neglect beam effects, which is reasonable since we are only interested in the largest cosmological scales. The noise between frequency bins $i$ and $j$ should be uncorrelated, so that the general expression for the instrumental noise angular power spectrum is given by
\be \label{eq:IMnoisepower}
\mathcal N_{\ell,ij}=\sigma_i^2\ \Omega_{{\rm pixel}, i}\ \delta_{ij}\,.
\ee

\subsection{\textsc{Hi} IM with an SKA1-like survey}

The redshift distribution of sources is given by the average \textsc{Hi} temperature, $p^{\rm \textsc{Hi}}(z)\propto T^{\rm \textsc{Hi}}(z)$. (For a detailed discussion on separating $T^{\rm \textsc{Hi}}(z)$ from \textsc{Hi} bias and other cosmological parameters in the measured power spectrum, see \citet{Bull:2014rha}.) We follow \citet{Santos:2015bsa} to compute both the Gaussian bias $b^{\rm \textsc{Hi}}_{\rm G}\z$ (their Eq. 2.5) and  temperature (their Eq. 2.1). Figure \ref{fig:bias} shows the \textsc{Hi} Gaussian bias as a function of redshift (thin solid blue).  We can write the comoving number density of \textsc{Hi} atoms in terms of the temperature: $n^{\rm \textsc{Hi}}\z\propto T^{\rm \textsc{Hi}}(z) H\z /(1+z)^2$. Then the evolution bias \eqref{be} becomes
\be 
b_e^{\rm \textsc{Hi}}(z)=-\frac{\p\ln \big[T^{\rm \textsc{Hi}}(z) H\z\big] }{\p\ln (1+z)}-2\,.
\ee
The experimental noise angular power spectrum in a redshift bin $i$ for an \textsc{Hi} IM experiment in single-dish mode is given by
\be
\mathcal N^{\rm \textsc{Hi}}_{ij}=\frac{4\pi \fsky\, T^2_{{\rm sys},i}}{2N_d\, t_{\rm tot}\,\Delta\nu^{\textsc{Hi}}_i}\ \delta_{ij}\,,\label{eq:N_HI}
\ee
where $T_{{\rm sys},i}$ is the system temperature for bin $i$, $\D\nu_i^{\textsc{Hi}}$ is the frequency width of the bin for \textsc{Hi}, $N_d$ the number of collecting dishes, $t_{\rm tot}$ the total observation time and $\fsky$ the observed fraction of the sky (we take $\fsky=0.75$). Note that  (\ref{eq:N_HI}) follows from  (\ref{eq:IMnoisepower}), since
\bea
\sigma^{\rm \textsc{Hi}}_i&=&\frac{T_{{\rm sys},i}}{\big(2 N_d\, \Delta\nu^{\textsc{Hi}}_i\, t_{{\rm pixel},i}\big)^{1/2} }\,,\\
\Omega^{\rm SKA}_{{\rm pixel},i}&=&\frac{\Omega_{\rm survey}}{N_{{\rm pixel},i}} = \frac{4\pi \fsky}{t_{\rm tot}/t_{{\rm pixel},i}}\,,
\eea
where $t_{{\rm pixel},i}$ is the integration time per pixel at frequency bin $i$, $N_{{\rm pixel},i}$ is the number of pixels in slice $i$ and $\Omega_{\rm survey}$ is the total surveyed area. 
We take $T_{\rm sys}=T_{\rm rec}+T_{\rm sky}$, where the receiver temperature is $T_{\rm rec}=25$ K and the sky contribution is given by $T_{\rm sky}=60\times (300\,{\rm MHz}/\nu)^{2.55}$ K \citep[see caption of Table 2 in][]{Santos:2015bsa}. We also assumed that $N_d t_{\rm tot}=2\times10^6\,$hr. 

\subsection{H$\alpha$ IM with a SPHEREx-like survey}

\begin{figure}
\centering
\includegraphics[width=\columnwidth]{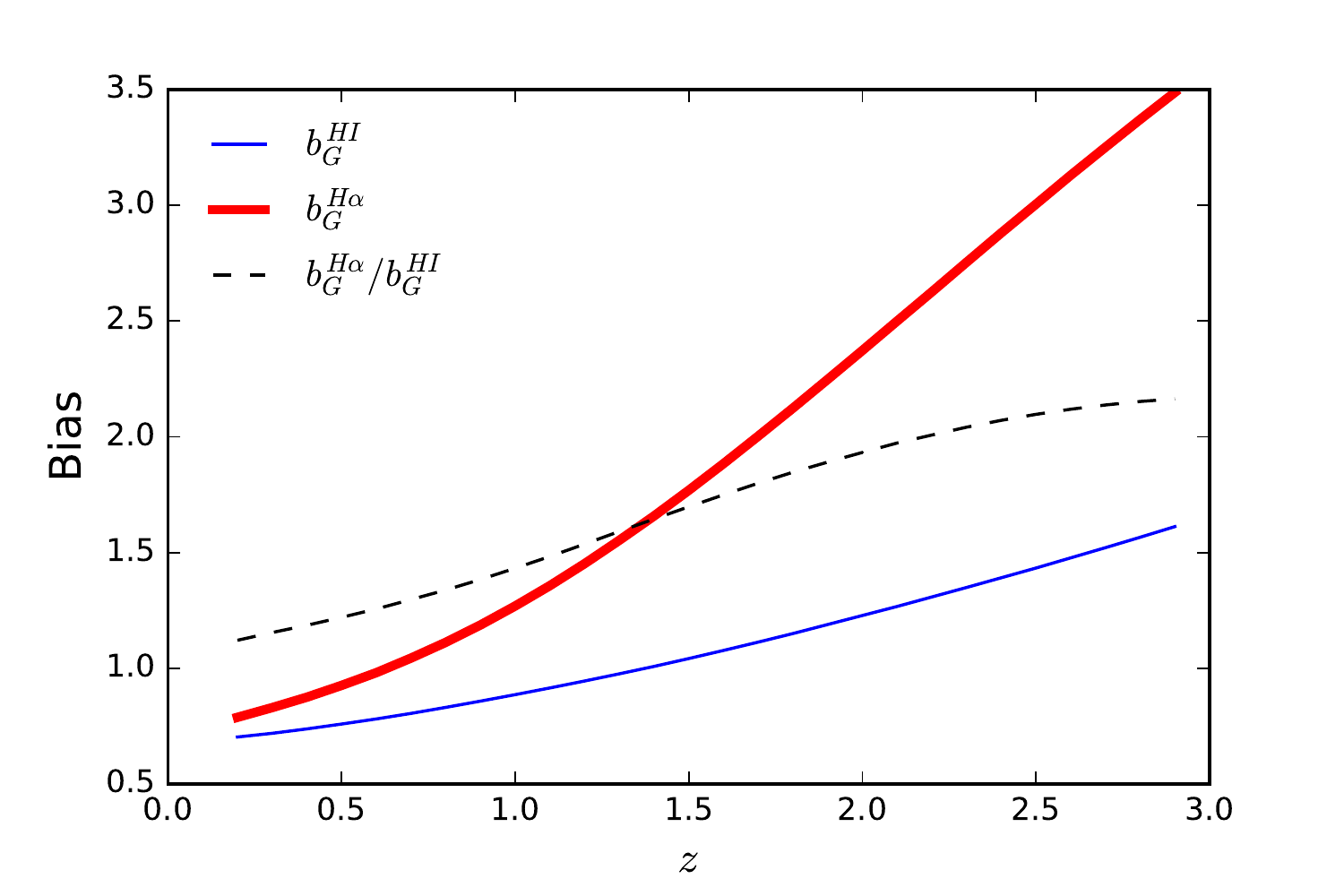}
\caption{Gaussian bias of  \textsc{Hi} \citep[thin solid blue]{Santos:2015bsa} and H$\alpha$ \citep[thick solid red]{Fonseca:2016qqw} emission lines. The ratio of the biases is also shown (dashed).}\label{fig:bias}
\end{figure}

The most energetic line of the Balmer series, H$\alpha$, with a rest-frame wavelength of 656\,nm, is also one of the strongest emission lines from galaxies. The emission comes from recombination of hydrogen atoms that have been ionised by young UV emitting stars. The H$\alpha$ line is therefore a tracer of star-forming galaxies \citep{Laureijs:2011gra,Amendola:2012ys,Spergel:2015sza}. The linear Gaussian bias  $b^{\rm H\alpha}_{\rm G}\z$ for H$\alpha$ IM is given by \citet{Fonseca:2016qqw} and shown in Fig. \ref{fig:bias} (thick solid red line). The H$\alpha$ bias is consistent with the measurement $b_G^{{\rm H}\alpha}=2.4^{+0.1}_{-0.2}$ at $z=2.23$ of \citet{2012MNRAS.426..679G}.

The average line intensity computed in \citet{Fonseca:2016qqw} is well described by the fit
\be \label{eq:IHa}
I_\nu^{\rm H\alpha}\z=\frac{7.79\times 10^{-15}(1+z)^{1.17}}{1+1.37\times 10^{-3}(1+z)^{6.61}}\ {\rm erg/s/cm^2/Hz/Sr}\,.
\ee
For H$\alpha$ IM the redshift distribution of sources is given by the intensity, i.e., $p^{\rm H\alpha}(z)\propto I_\nu^{\rm H\alpha}(z)$. We assume that $b^{\rm H\alpha}_{\rm e}\z\simeq b^{\rm \textsc{Hi}}_{\rm e}\z$, since the comoving number density of hydrogen atoms is independent of hydrogen emission lines. This follows from assuming that the number density of H$\alpha$ emitters is a constant fraction (or with a negligible redshift evolution) of the average number of neutral hydrogen atoms.

The proposed space telescope SPHEREx would have a spectral resolution of $\la/\Delta\la= 41.5$ for $0.75<\la<4.1\,\mu$m. Although it is mainly intended as a galaxy survey, it can have a H$\alpha$ IM mode. 
The proposed instrument has a pixel size $\Omega^{\rm SPHEREx}_{\rm pixel}=6.2''\times6.2''\sim 10^{-9}\,$sr. We assume a constant $1\sigma$ flux sensitivity of $\sigma_F\sim 10^{-17}$ erg/s/cm$^2$, higher than the SPHEREx $5\sigma$ sensitivity of 18.4 AB magnitudes \citep{Dore:2014cca}. According to \citet{Silva:2017aa} (their Fig. 1), \eqref{eq:IHa} may under-estimate the average H$\alpha$ intensity. The results are sensitive to the ratio $I_\nu^{H\alpha}/\sigma( {I_\nu^{H\alpha}})$ and not to the sensitivity or the intensity alone. For this reason, we choose to increase flux sensitivity instead of increasing the intensity of the signal. The flux sensitivity translates to an uncertainty in the intensity given by $\sigma(I_\nu^{\rm H\alpha})=\sigma_F/\big(\Omega^{\rm SPHEREx}_{\rm pixel}\, \delta\nu\big)$, for the experimental angular resolution $\Omega^{\rm SPHEREx}_{\rm pixel}$ and the experimental frequency resolution $\delta\nu$ at observed frequency $\nu$. We use survey strategies with thicker frequency bins than the resolution one. This reduces the error in determining the intensity by $\sigma_i(I_\nu^{\rm H\alpha})=\sigma(I_\nu^{\rm H\alpha})\,\big(N_{\nu,i}\big)^{-1/2}$, where the number of resolution bins in a survey bin is $N_{\nu,i}=\Delta\nu_i^{\rm H\alpha}/\delta\nu$. Note that $\Delta\nu_i^{\rm H\alpha}$ is the frequency width of the bin for H$\alpha$. The noise angular power spectrum is then given by \eqref{eq:IMnoisepower} as
\bea
{\mathcal N}_{ij}^{\rm H\alpha}&=& \left[\sigma_i\big(I_\nu^{\rm H\alpha}\big)\right]^2\, \Omega^{\rm SPHEREx}_{\rm pixel}\ \delta_{ij}  \,,\nn\\
&=&\left(\frac{\sigma_F}{\Omega^{\rm SPHEREx}_{\rm pixel}\, \delta\nu_i}\right)^2\,\frac{\delta\nu_i}{\Delta\nu_i^{\rm H\alpha}}\, \Omega^{\rm SPHEREx}_{\rm pixel}\ \delta_{ij}\,,\label{eq:N_Ha}
\eea 
where $\delta\nu_i$ is the frequency resolution of SPHEREx at the central frequency of bin i.

\section{Fisher Forecast Analysis} \label{sec:fisher_forecasts}

The Fisher matrix for a set of parameters $\vartheta_i$ is 
\bea \label{eq:fishergeneral} 
{ F}_{\vartheta_i\vartheta_j}=\frac12 {\rm Tr}\left[ \left(\partial_{\vartheta_i}{ C}\right) {\Gamma}^{-1}\left(\partial_{\vartheta_j} {  C}\right){ \Gamma}^{-1}\right], ~~~{ \Gamma}={  C}+{  \cal N},
\eea
where ${C}$ is the covariance matrix of the estimator and ${ \cal N}$ is the noise term. For IM, shot-noise is subdominant to instrumental noise, so we can safely assume that the noise term is independent of the $\vartheta_i$. Since the observables are the $a^{AB}_{\ell m}$, the estimator's covariance is the angular power spectrum $C^{AB}_\ell(z_i,z_j)$. Considering the information in a given range of multipoles, we rewrite \eqref{eq:fishergeneral} as \citep{Tegmark:1996bz}
\bea \label{eq:fishercl}
{ F}_{\vartheta_i\vartheta_j}=\sum_{\ell_{\rm min}}^{\ell_{\rm max}} \frac{(2\ell+1)}{2 }\fsky\,{\rm Tr}\left[ \left(\partial_{\vartheta_i}{  C}_{\ell}\right){ \Gamma}_\ell^{-1} \left(\partial_{\vartheta_j}{  C}_{\ell}\right){ \Gamma}_\ell^{-1}\right]\,.
\eea
The sky fraction $\fsky$ enters via an approximation that accounts for the fact that not all $m$ are accessible at a given $\ell$. For the MT technique one requires that the sky maps of the differently biased tracers cover the same sky area and have the same redshift slicing. Then the covariance matrix is \citep{Ferramacho:2014pua} 
\bea
{ C}^{AB}_\ell\l z_i,z_j\r
=
\begin{pmatrix}
    {  C}^{\textsc{Hi},\textsc{Hi}}_{\ell,ij} & { C}^{\textsc{Hi},\rm H\alpha}_{\ell,ij} \\ &&\\
   { C}^{\rm H\alpha,\textsc{Hi}}_{\ell,ij} &  { C}^{\rm H\alpha, H\alpha}_{\ell, ij} 
\end{pmatrix}\,,
\eea
where $i,j$ indicate the redshift bins. Assuming that we are dealing with Gaussian likelihoods, the inverse of the Fisher matrix is a good approximation to the parameter covariance. In this case the forecast marginal error for  parameter $\vartheta_i$ is 
\bea
\sigma_{\vartheta_i}=\big[{({  F}^{-1})_{\vartheta_i\vartheta_i}}\big]^{1/2}\,.
\eea

We consider the set of parameters
\bea \label{para}
  \vartheta_\alpha &=& \Big\{ \ln A_s,\ln n_s,\ln\odm,\ln \ob,w, \ln H_0,  \nn\\
  &&~~~ b^{\rm \textsc{Hi}}_G(z_i), b_G^{\rm H\alpha}(z_i), \fnl, 
   \varepsilon_{\rm Doppler},\varepsilon_{\rm SW},\varepsilon_{\rm ISW}\,   \Big\}.
\eea  
For the cosmological parameters we used the fiducial values: $A_s=2.142\times 10^{-9}$, $n_s=0.967$, $\odm=0.26$, $\Omega_{b}=0.05$, $w=-1$, $H_0=67.74\,$ km/s/Mpc and $\fnl=0$. The fiducial values for the Gaussian clustering biases were discussed in \S \ref{sec:surveys} and are shown in Fig. \ref{fig:bias}. The parameters  $\varepsilon_{\rm Doppler}$, $\varepsilon_{\rm SW}$ and $\varepsilon_{\rm ISW}$ are introduced to assess how well we can measure each individual contribution of the GR effects \eqref{dop2}--\eqref{isw2}. They have a fiducial value of $\varepsilon=1$, and are intrinsically defined by
\be \label{eps_def}
\D_\ell^A=\D_\ell^{A\,{\rm standard}}+\varepsilon_{\rm Doppler} \D_\ell^{A\,{\rm Doppler}}+\varepsilon_{\rm SW} \D_\ell^{A\,{\rm SW}}+\varepsilon_{\rm ISW} \D_\ell^{A\,{\rm ISW}}\,.
\ee

To compute the angular power spectra and their derivatives with respect to the cosmological parameters we used a modified version\footnote{\url{https://github.com/ZeFon/CAMB\_sources\_MT\_ZF.git}} of \cs. The derivatives with respect to $\ln\odm,\ln \ob,w, \ln H_0$ are evaluated numerically using the 5-point stencil method. For the remaining parameters, they are done analytically (see Appendix \ref{appder}).  The supporting Python wrappers to compute the derivatives and the Jupyter Notebook used to compute the Fisher matrix and results presented in this paper can be found in the GitHub repository\footnote{\url{https://github.com/ZeFon/mt\_ska\_spherex.git}}.

We assume a redshift range $0.2\leq z\leq 3$ and we compare three binning strategies to see their effects on forecasts:  conservative, with 14 thick bins of width $\D z=0.2$; less conservative, with 28 bins of size $\D z=0.1$; using the full spectral resolution of SPHEREx, i.e., $\D z_{\rm min}= (1+z)/41.5$. SKA1 has much higher redshift resolution, but the MT requires equal binning. 

The Fisher matrix depends on the surveyed area either directly via $\fsky=\Omega_{\rm survey}/4\pi$  in \eqref{eq:fishercl}, or indirectly via the maximum accessible scale, $\ell_{\rm min}= 1+{\rm integer}\,(\pi/\sqrt{\Omega_{\rm survey}})$. We  assume an overlap sky fraction of $f_{\rm sky}=0.75$, so that $\ell_{\rm min}=2$. 

Our focus is on ultra-large scale effects, so we do not need to model nonlinear scales, $k>k_{\rm NL}$, where \citep{Smith:2002dz}
\bea
k_{\rm NL}(z)&\approx&k_{\rm NL,0}\l1+z\r^{2/(2+n_s)},\quad k_{\rm NL,0}\approx 0.2 h\,{\rm Mpc}^{-1}\,,\\ \ell_{\rm NL}&\approx & \chi k_{\rm NL}\,,
\eea
and we used the Limber approximation. We  cap contributions to the Fisher matrix from a redshift bin \emph{i} at $\ell^i_{\rm max}={\rm max}\,(300,\ell^i_{\rm NL})$. This is only relevant for the lowest redshift bins where  $\ell_{\rm NL}$ is small. At high $z$, $\ell_{\rm max}=300$ is conservative, but this has a negligible effect on our results. 

\section{Results} \label{sec:results}

\begin{table}
\caption{\label{tab:marginal}Marginal errors on $\fnl$, $\varepsilon_{\rm Doppler}$, $\varepsilon_{\rm SW}$ and $\varepsilon_{\rm ISW}$ for different binnings $\Delta z$, for the two IM surveys separately and combined via MT analysis, assuming $f_{\rm sky}=0.75$, $0.2\leq z\leq 3$, 
and the instrumental noise given in \S \ref{sec:surveys}.}
\centering
\begin{tabular}{lrcccc}
$\Delta z$&&$\s(\fnl)$ & $\s(\varepsilon_{\rm Doppler})$ & $\s(\varepsilon_{\rm SW})$& $\s(\varepsilon_{\rm ISW})$\\
\hline
&\textsc{Hi}&5.0&5.9&14&11\\
0.2&\textsc{H$\alpha$}&3.9&8.1&30&22\\
&MT&1.2&0.64&6.0&9.7\\
\hline
&\textsc{Hi}&4.3&4.9&12&11\\
0.1&\textsc{H$\alpha$}&3.5&6.8&27&22\\
&MT&1.1&0.39&5.6&9.5\\
\hline
&\textsc{Hi}&3.5&4.3&11&9.5\\
$\Delta z_{\rm min}$&\textsc{H$\alpha$}&3.4&6.3&26&20\\
&MT&1.0&0.33&4.9&8.3\\
\hline
\end{tabular}
\end{table}

The forecast marginal errors for the ultra-large scale parameters $\fnl$,  $\varepsilon_{\rm Doppler}$, $\varepsilon_{\rm SW}$ and $\varepsilon_{\rm ISW}$ are summarised in Table \ref{tab:marginal}. Results are shown for each binning strategy. For comparison,  the single-tracer constraints with the same binning are shown. Increasing the redshift resolution does improve the results mildly. (Note that the $\fnl$ constraints for SKA1 would improve considerably if the greater available spectral resolution were used.) As expected, the MT technique improves the errors, especially for the Doppler term. The forecast $\sim3\sigma$ detection of the Doppler term is consistent with the results of \citet{Abramo:2017xnp}, who applied the MT technique to two galaxy surveys, based on the monopole and dipole of the Cartesian power spectrum. We marginalised over all the other parameters, since our focus is on $\fnl$ and the GR effects. Marginalisation includes any correlations between $A_s$ and the $\varepsilon$ parameters in the quoted errors. Since $A_s$ is an overall scale-independent amplitude, while $\fnl$ and the $\varepsilon$ parameters are multiplied by terms with $1/k$ or $1/k^2$ dependences, we expect little correlation with $A_s$.

\begin{table}
\caption{\label{tab:conditional} As in Table \ref{tab:marginal}, for conditional errors on $\fnl$.}
\centering
\begin{tabular}{lccc}
$\Delta z$&\textsc{Hi}&\textsc{H$\alpha$}& MT\\
\hline
0.2 & 2.4 & 1.4 & 1.0\\
0.1 & 2.1 & 1.3 & 0.92\\
$\Delta z_{\rm min}$ & 1.8 & 1.3 & 0.86\\
\hline
\end{tabular}
\end{table}

In Table \ref{tab:conditional}, for comparison we give the conditional errors on $\fnl$. For the single-tracer case there is a substantial improvement from marginal to conditional error, as expected. This is not true for the multi-tracer case: the auto- and cross-correlations reduce degeneracies between parameters, leading to a smaller improvement.
The single-tracer case does not give prospects of crossing the $\sigma_{\fnl}\sim1$ threshold, due to cosmic variance, in agreement with previous results \citep{Camera:2013kpa,Camera:2014bwa,Alonso:2015uua,Raccanelli:2015vla}. The MT results just reach $\sigma_{\fnl}\approx 1$ (marginal error), not as good as MT with \textsc{Hi} IM and photometric surveys \citep{Alonso:2015sfa,Fonseca:2015laa}. The MT technique is known to improve with the increasing difference in bias-like parameters of the tracers. For two IM surveys, the clustering biases have a small ratio (see Fig. \ref{fig:bias}), the evolution biases are equal, and both have no magnification bias. By contrast, an IM and a photo-$z$ survey have significant differences in all 3 astrophysical parameters \citep{Alonso:2015sfa,Fonseca:2015laa}.

\begin{figure}
\centering
\includegraphics[width=\columnwidth]{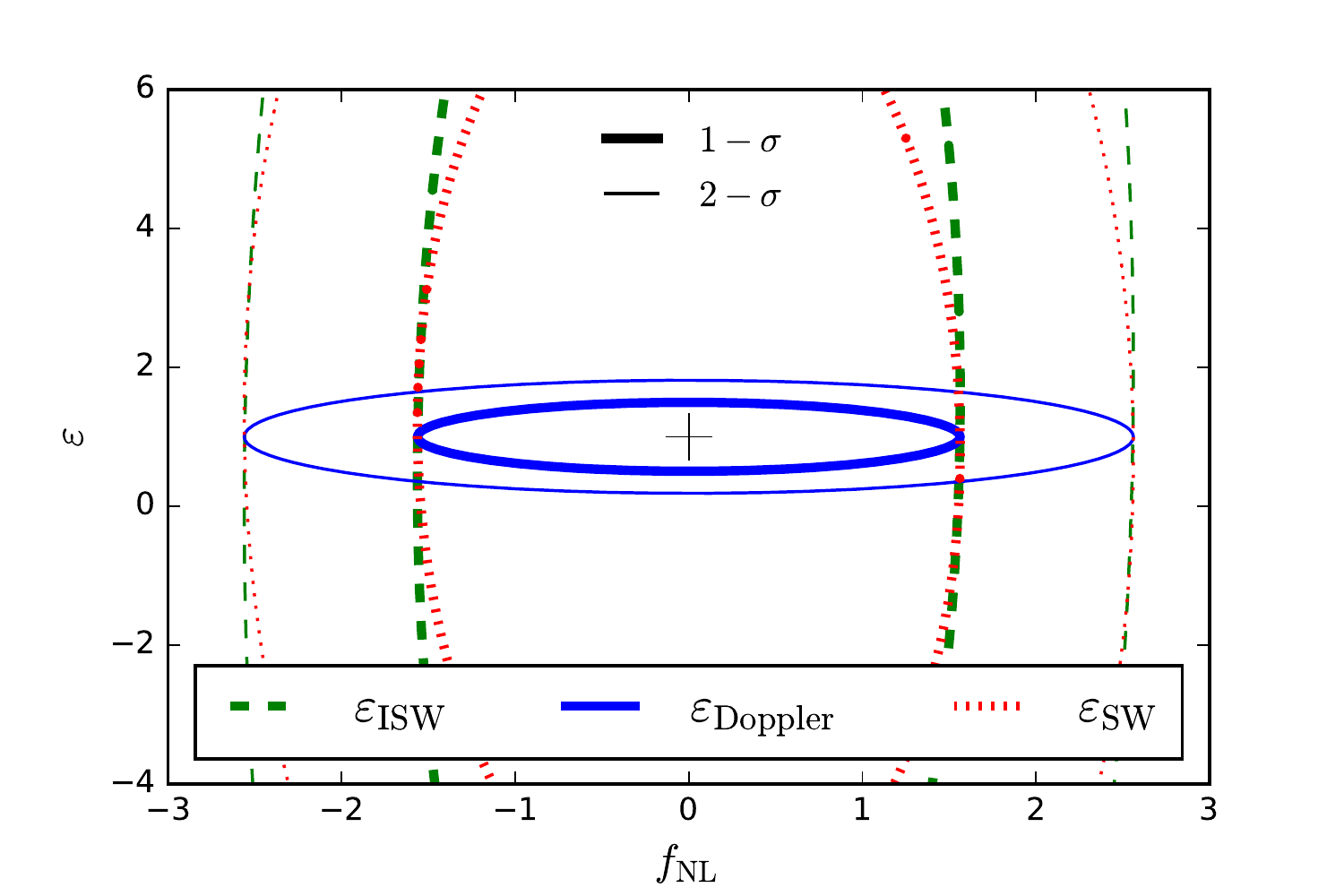}
\caption{The $1\sigma$ (thick) and $2\sigma$ (thin) contours for the marginal errors of $\fnl$ and the Doppler (solid blue), SW (dashed red) and ISW (dot-dashed green) GR  effects. ($\fsky=0.75$, $0.2\leq z\leq 3$ and redshift resolution $\Delta z_{\rm min}$.) }\label{fig:onsesigma_fnl_eps}
\end{figure}

Constraints on $\fnl$ are intimately connected to those on the GR effects, given the partial degeneracy between the effects of primordial non-Gaussianity and those of GR corrections \citep{Bruni:2011ta,Jeong:2011as,Camera:2014sba}.
This can be seen in Fig. \ref{fig:onsesigma_fnl_eps}, showing the forecast $1\s$ and $2\s$ marginalised contours for $\fnl$ with the different GR effects.  
Table \ref{tab:marginal} shows that the Doppler term is well constrained by the IM MT (with maximal redshift resolution), implying that it would be detectable at $\sim3\sigma$. Neither the SW nor the ISW terms are constrained by the MT.

\begin{figure}
\centering
\includegraphics[width=\columnwidth]{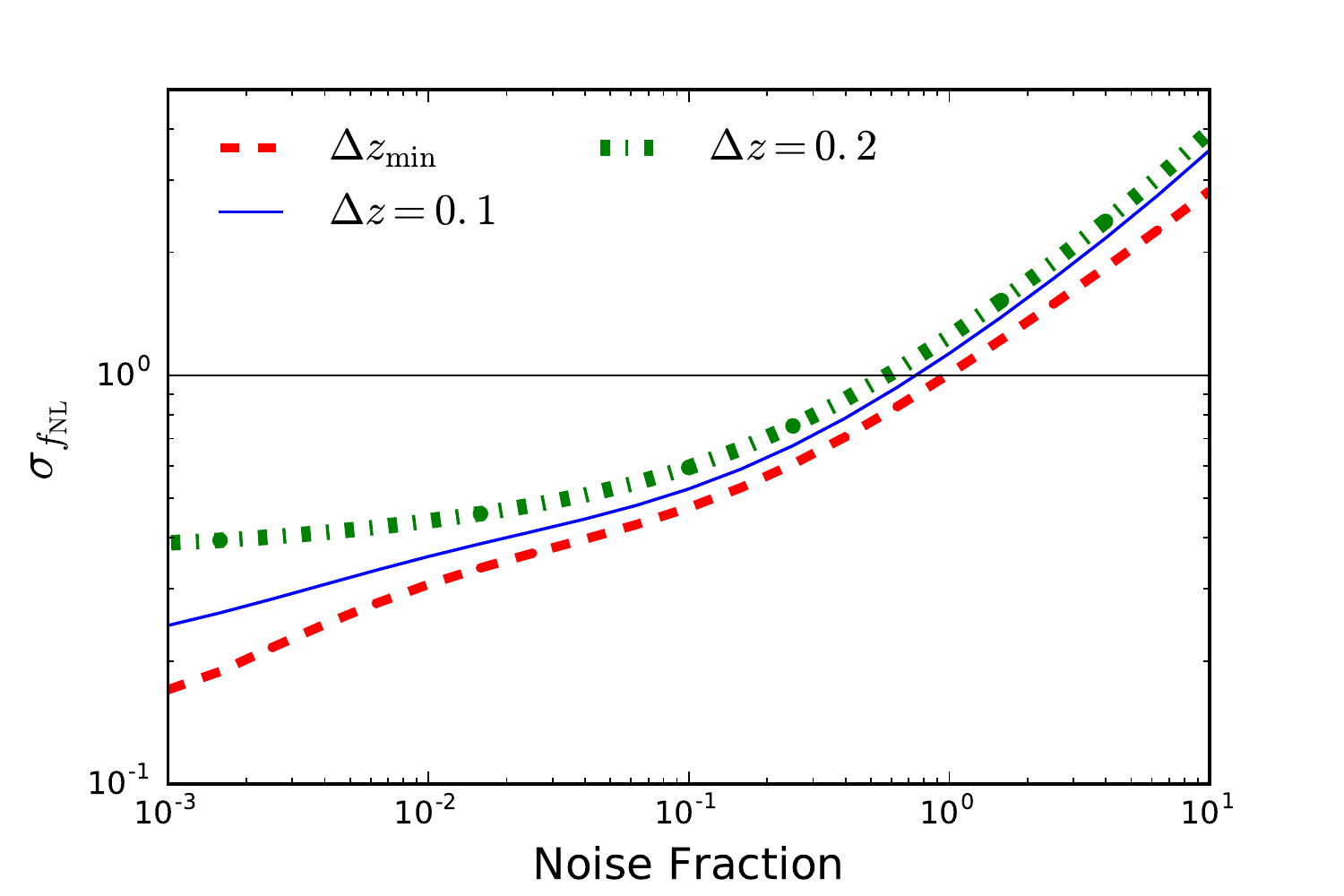}
\caption{Dependence of the forecast $\sigma(\fnl)$ on the noise, as a fraction of the reference noise ($=1$), for the three binnings: $\Delta z=0.2$ (thick green dot-dashed), $=0.1$ (thin blue solid), $=\Delta z_{\rm min}$  (red dashed). The horizontal line indicates the threshold $\sigma(\fnl)=1$. ($\fsky=0.75$, $0.2\leq z\leq 3$.)}\label{fig:noise_dependence_fnl}
\end{figure}

\begin{figure}
\centering
\includegraphics[width=\columnwidth]{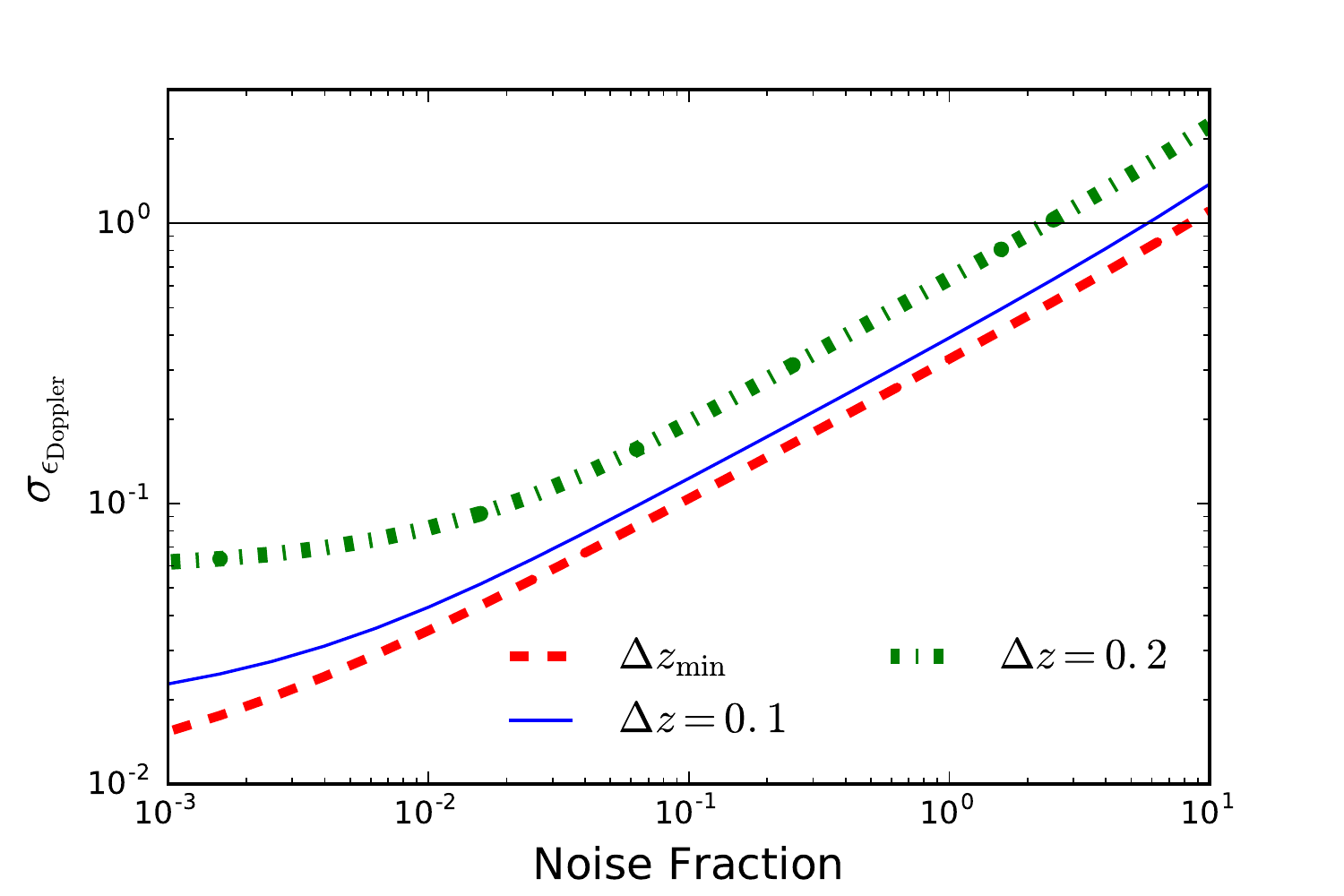}
\includegraphics[width=\columnwidth]{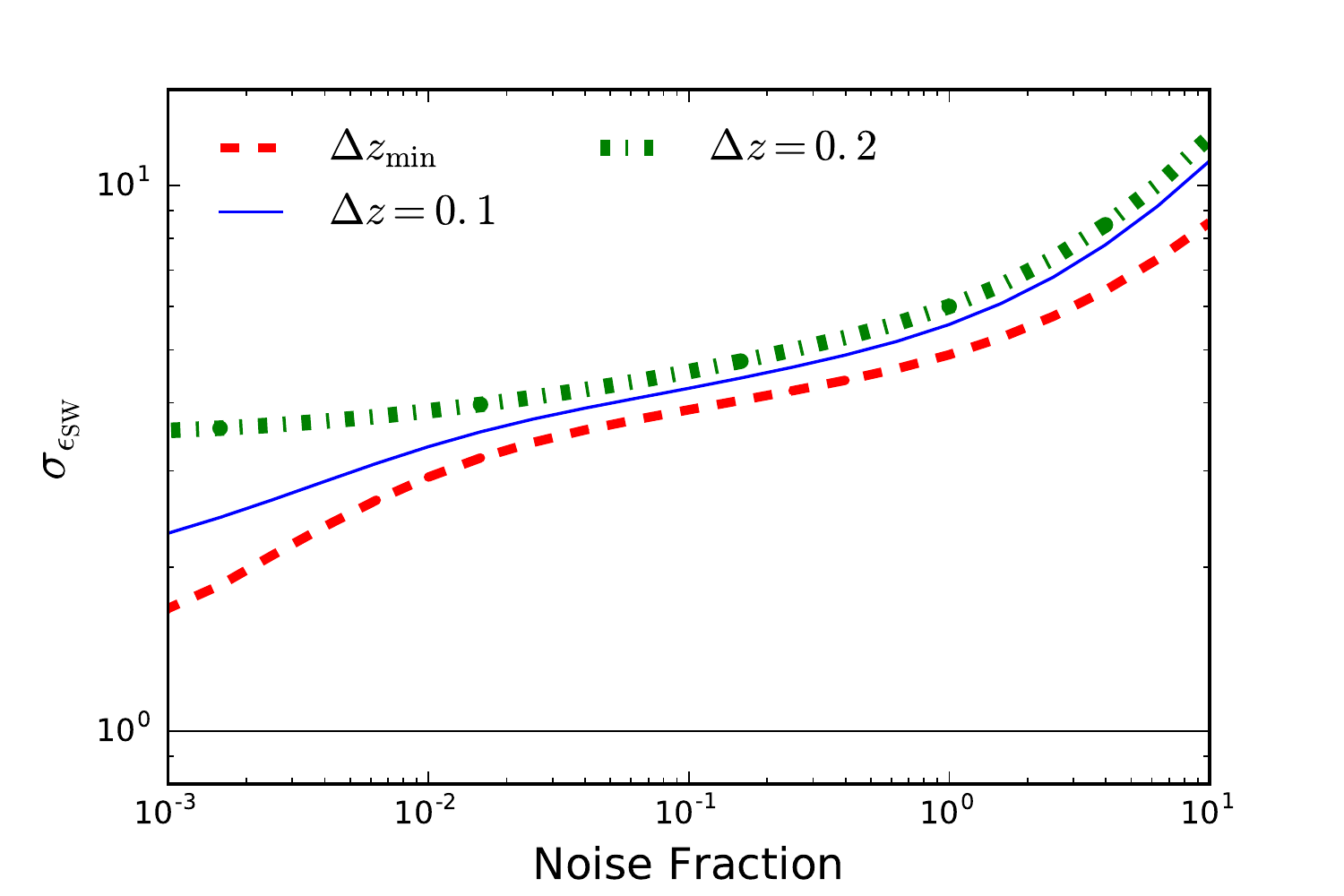}
\includegraphics[width=\columnwidth]{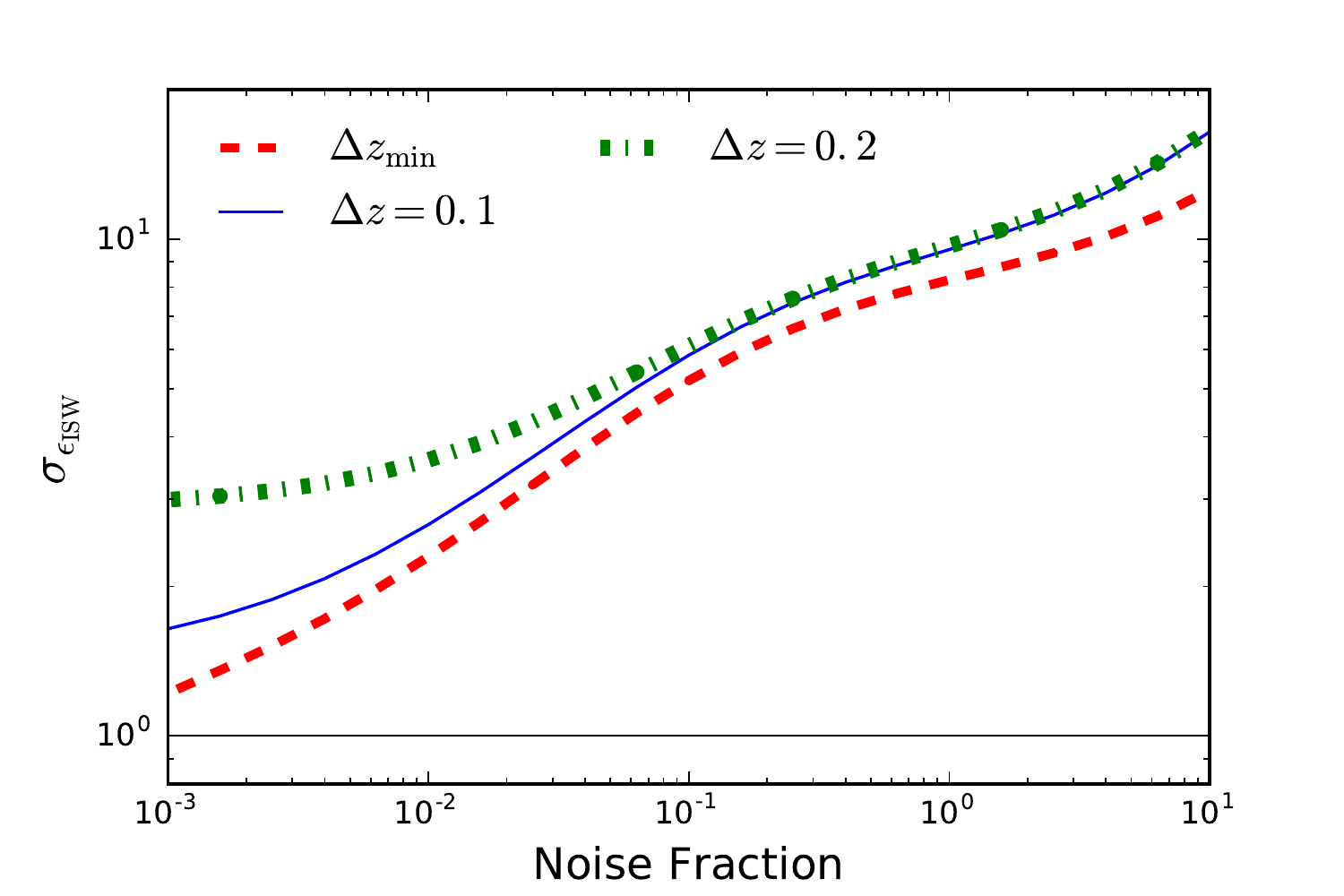}

\caption{As in Fig. \ref{fig:noise_dependence_fnl}, for the GR parameters:  $\varepsilon_{\rm Doppler}$ ({\em top}), $\varepsilon_{\rm SW}$ ({\em middle}), $\varepsilon_{\rm ISW}$ ({\em bottom}).}\label{fig:noise_dependence_gr}
\end{figure}

With the MT we are no longer cosmic variance limited. The instrumental noise is thus the most important source of uncertainty for ultra-large scale parameters. We see this in Figs. \ref{fig:noise_dependence_fnl} and \ref{fig:noise_dependence_gr}, showing the forecast errors as a function of the fraction of the reference noise ($=1$), for the three different binning strategies. While the errors are sensitive to the redshift resolution, they are even more sensitive to the experimental noise. The reference noise is given in \S \ref{sec:surveys} and we vary the noise simultaneously for the two experiments, allowing also for higher noise. While for SKA1 one can lower the noise by increasing the observation time, one would need to wait for a more futuristic H$\alpha$ experiment to improve sensitivity.

It is also interesting to compare how this combination of tracers would  perform if we bundle all GR effects together, as in \cite{Alonso:2015sfa} and \cite{Fonseca:2015laa}. The forecast result using the MT technique ranges from $\sigma_{\varepsilon_{\rm GR}}(\delta z=0.2)=0.63$ to $\sigma_{\varepsilon_{\rm GR}}(\delta z_{\rm min})=0.33$. This is consistent with Table \ref{tab:marginal}, i.e. detection of the collective GR effects is  basically detection of the Doppler effect for the MT of the two intensity maps.

\section{Discussion} \label{sec:discussion}

\begin{figure}
\centering
\includegraphics[width=\columnwidth]{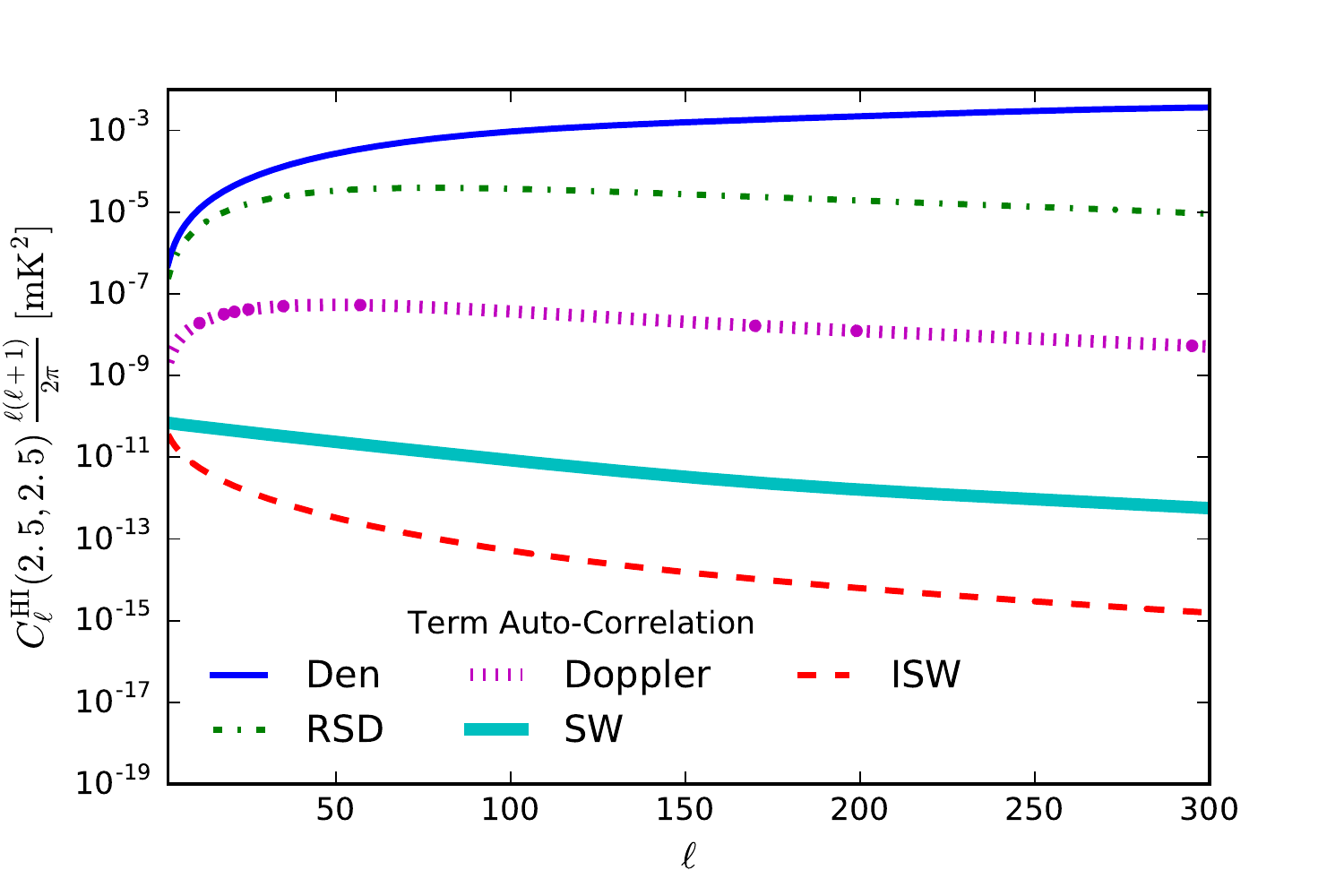}
\caption{Angular power auto-spectra $C_\ell(2.5,2.5)$ at $z=2.5$ of each term in Eq. \eqref{eq:angIM}: density (thin solid blue); RSD (thin dot-dashed green); Doppler (thick dotted magenta); SW (thick solid cyan); ISW (thin dashed red). We use a Gaussian window function with $\sigma_z=0.05$.}\label{fig:auto_contribution}
\end{figure}
\begin{figure}
\centering
\includegraphics[width=\columnwidth]{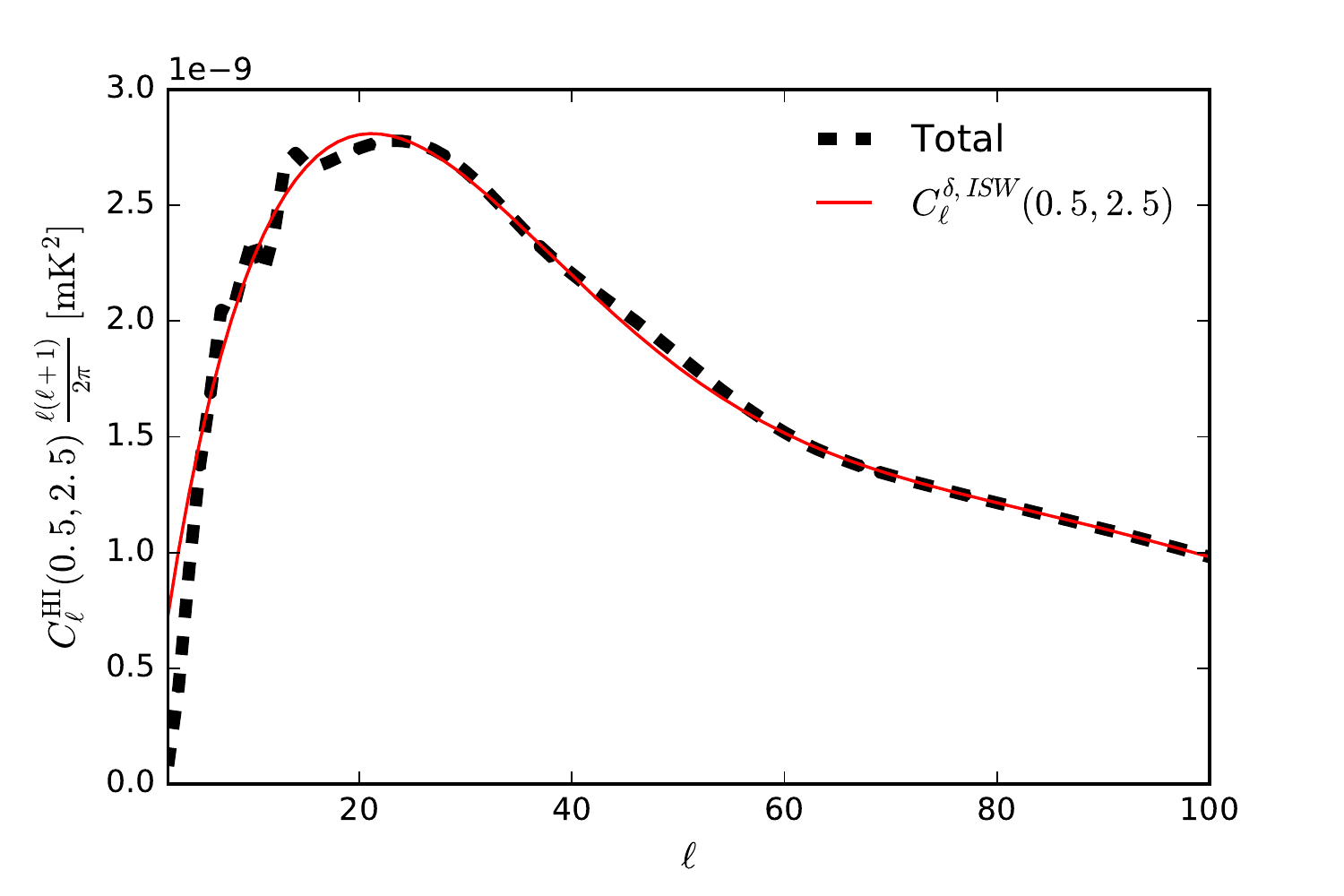}
\caption{Angular power cross-spectra $C_\ell(0.5,2.5)$: total spectrum (dashed black); density ($z=0.5$)--ISW ($z=2.5$) (solid red). }\label{fig:crossbin}
\end{figure}

We have shown that a MT combination of intensity maps of two hydrogen lines can detect the GR Doppler contribution to intensity, but the remaining GR effects, SW and integrated SW, are not detectable. In the best case scenario, the Doppler term is constrained at $3\sigma$. 
To our knowledge, this is the first time that the GR effects have been separately constrained.  (The GR effect of weak lensing convergence has been separately constrained, but this effect vanishes at first order for intensity mapping.)

By Eqs. \eqref{dop2}--\eqref{isw2}, the Doppler term scales as $v_k \propto {\cal H}/k$, while the SW and ISW terms scale as $\phi_k\propto {\cal H}^2/k^2$. The Doppler effect is thus stronger for $k>{\cal H}$, and is the dominant GR contribution to the auto-correlations of each intensity map. This is illustrated in Fig. \ref{fig:auto_contribution}.
For cross-bin correlations, we expect that the ISW should dominate for widely separated bins. This is confirmed by Fig. \ref {fig:crossbin}, which shows that the angular power cross-spectrum between bins at $z=0.5$ and  $z=2.5$ is dominated by the low-$z$ density correlation with high-$z$ ISW.

We also showed that
the same MT combination can constrain the primordial non-Gaussianity parameter down to the level $\sigma(\fnl)\sim 1$.
In contrast to other works where the MT has been used to break through the $\sigma_{\fnl}\sim1$ threshold \citep[see eg][]{Ferramacho:2014pua, Yamauchi:2014ioa,Dore:2014cca,Alonso:2015sfa,Fonseca:2015laa}, we used two intensity maps rather than combinations of galaxy surveys or a galaxy survey with \textsc{Hi} IM. The combination of maps of intensity of two hydrogen lines is  not optimal to measure the local non-Gaussian parameter $\fnl$, since the clustering biases are both close to one.

We can explain the difference in detectability amongst the GR terms as follows. The observed fluctuations satisfy
\bea
\D^{\rm H\alpha}-\D^{\textsc{Hi}}=\l b^{\rm H\alpha}- b^{\textsc{Hi}}\r\delta^s\,, \label{eq:difD}
\eea
since the evolution biases are equal, $b^{\textsc{Hi}}_e=b^{\rm H\alpha}_e$. From this we can find a combination that excludes the density constrast:
\bea
\D^{\rm H\alpha}-\frac{b^{\rm H\alpha}}{b^{\textsc{Hi}}}\D^{\textsc{Hi}}=\l1-\frac{b^{\rm H\alpha}}{b^{\textsc{Hi}}}\r\bigg[\D^{\rm RSD}+\D^{\rm Doppler} + \D^{\rm SW}+\D^{\rm ISW} \bigg]\label{eq:difDvel}\,.
\eea
While the difference of the fluctuations is only  sensitive to the clustering bias (and thus primordial non-Gaussianity), the second combination is mainly sensitive to velocity terms -- RSD and Doppler. The RSD contribution is the dominant term, but the Doppler contribution correlates perfectly with RSD, thus enhancing the chances of detecting this GR term. By the same token, the fact that we still have RSD in the second combination makes it the dominant source of intrinsic uncertainty if we want to measure the SW and ISW in the large-scale structure. The second combination therefore illustrates why the Doppler term is detectable while the SW and ISW terms are not. The latter terms are overwhelmed by the uncertainty in RSD.
If we neglected RSD and Doppler effects, i.e. if we assumed that density and SW/ ISW were the only contributions, then the second combination of fluctuations implies that SW and ISW should be detectable.

\begin{figure}
\centering
\includegraphics[width=\columnwidth]{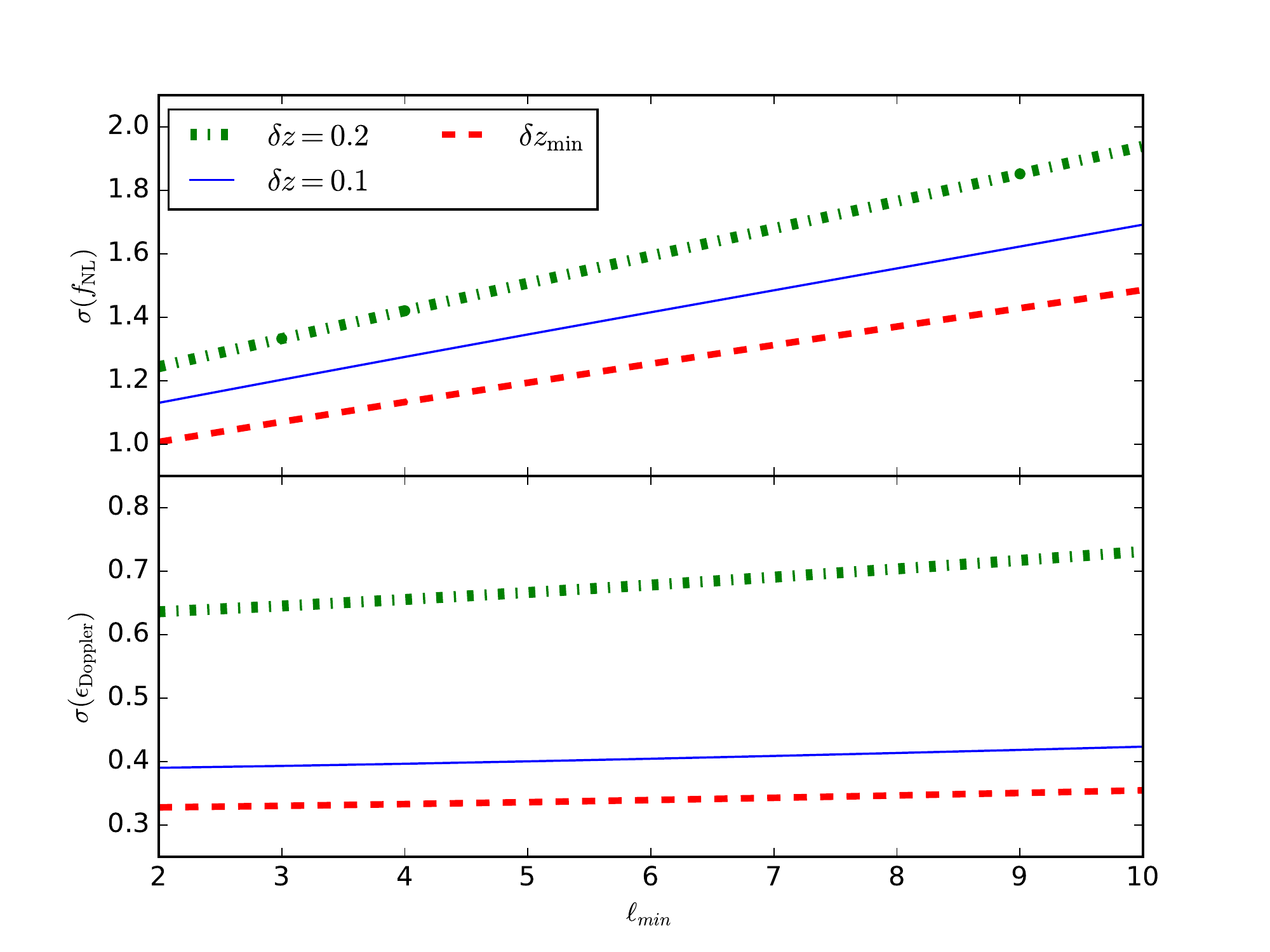}
\caption{Forecast errors in measuring $\fnl$ (top) and the Doppler term (bottom) as a function of the largest included scale $\ell_{\rm min}$, for the three chosen binning strategies.}\label{fig:sig_vs_ellmin}
\end{figure}

We have assumed that the foregrounds for the two intensity maps have been dealt with using well-developed cleaning techniques \citep[see, e.g.][for \textsc{Hi} and H$\alpha$, respectively]{Wolz:2015sqa,Silva:2017aa}. Note that any foreground residuals left after the cleaning/masking procedure will mainly affect $C_\ell^{\textsc{Hi,Hi}}$ and $C_\ell^{{\rm H}\alpha,{\rm H}\alpha}$, since in $C_\ell^{\textsc{Hi},{\rm H}\alpha}$ the foregrounds of \textsc{Hi} should not correlate with those  of H$\alpha$. In practice, such foreground removals will affect the minimum multipole $\ell_{\rm min}$ that we can use to compute parameter constraints. In Fig. \ref{fig:sig_vs_ellmin} we plot the dependence of $\sigma(\fnl)$ and $\sigma(\varepsilon_{\rm Doppler})$ as a function of $\ell_{\rm min}$. While the constraints on the Doppler term are not appreciably sensitive to $\ell_{\rm min}$, the error on $\fnl$ degrades more quickly as one discards the largest scales. \citet{Alonso:2014dhk} showed that \textsc{Hi} foreground removal mainly affects the largest modes in the line-of-sight (their Fig. 3 right panel). Despite this, the angular power spectrum can still be recovered on large scales (low $\ell$) but with a larger uncertainty (their Fig. 3 left panel). A more realistic approach should take these larger uncertainties at low $\ell$ into account, although it is beyond the scope of this paper. 

\subsection*{Acknowledgments}
We thank Stefano Camera, Marta Silva and an anonymous referee for useful comments. 
We are supported by the South African Square Kilometre Array Project and National Research Foundation. {We acknowledge the support of the Centre for High Performance Computing, South Africa, under the project ASTR0945.} RM is also supported by the UK Science \&\ Technology Facilities Council Grant No. ST/K0090X/1.

\appendix
\section{Analytical Derivatives of parameters} \label{appder}

To improve on numerics we take analytical derivatives whenever possible. Otherwise we computed the numerical derivatives using the 5-point stencil expression, 
\bea
\frac{{\rm d} C_\ell}{{\rm d} \theta}\simeq\frac{C_\ell(\theta-2h) - 8 C_\ell(\theta-h)+8 C_\ell(\theta+h)- C_\ell(\theta+2h)}{12h},
\eea
for a given interval $h$. This is done for $\ln\odm,\ln \ob,w, \ln H_0$.

For the parameters of the primordial power spectra: 
\bea
\frac{\partial C^{AB}_\ell(z_i,z_j)}{\partial A_s} &=& \frac{ C^{AB}_\ell(z_i,z_j)}{A_s}
\\\frac{\partial C^{AB}_\ell(z_i,z_j)}{\partial n_s} &=& 4\pi\!\!\int\!\!\d \ln k\,\D_\ell^{W_A}\!\l z_i,k\r\ \D_\ell^{W_B}\!\l z_j,k\r\ \,\ln\l\frac{k}{k_0}\ \r \mathcal P\l k\r \,.\nonumber\\&&
\eea

The derivatives with respect to the (Gaussian) clustering bias in each bin are
\begin{align}
 \frac{\partial C^{AB}_\ell(z_i,z_j)}{\partial b^C_G(z_n)}&\! = \! \int\!\! {\rm d}\ln k \bigg[\Delta^{W_B}_\ell(z_j)\ \delta_{ni}\ \delta^{AC} 
\! \int\!\! {\rm d}z' p^A(z') W(z',z_i)\ \delta_k j_\ell(k\chi) \nn\\
& \! +\Delta^{W_A}_\ell(z_i)\  \delta_{nj}\ \delta^{BC}
\int\!\! {\rm d}z'' p^B(z'') W(z'',z_j)\ \delta_k j_\ell(k\chi) \bigg] {\cal P}(k).\nn\\
\end{align}
Note that this is different than just asking for density in one window and making $b_n=1$, since the code is in Newtonian gauge. We change the code to compute each term in the last line.

The derivative with respect to $\fnl$ is
\bea
\frac{\partial C^{AB}_\ell(z_i,z_j)}{\partial \fnl}\! &=&\! \int \!\!{\rm d}\ln k \bigg[ \Delta^{W_B}_\ell(z_j) 
\!\int {\rm d}z' p^A(z') W(z',z_i) \frac{\partial \Delta b_i}{\partial f_{\rm NL}}\ \delta_k j_\ell(k\chi) \nn\\
&& \!+\Delta^{W_A}_\ell(z_i)
\!\int\!\! {\rm d}z'' p^B(z'') W(z'',z_j)  \frac{\partial \Delta b_j}{\partial f_{\rm NL}} \delta_k j_\ell(k\chi) \bigg] {\cal P}(k)\nn\\
&=& C^{AB}_\ell(z^{\delta \fnl}_i,z^{\rm all}_j) + C^{AB}_\ell(z^{\rm all}_i,z^{\delta \fnl}_j)\,,
\eea
with
\bea
\frac{\partial \Delta b_i}{\partial f_{\rm NL}}=3 \big[b_{\rm G}(z) -1\big]{D(z_{\rm dec})\left(1+z_{\rm dec}\right)}\frac{\Omega_m H_0^2 \delta_c}{D(z) T(k)  k^2}=\frac{\Delta b_i}{f_{\rm NL}}\,.
\eea

To identify if a GR correction to the angular power can be detected, we introduced the fudge factors in  \eqref{eps_def}, which we can rewrite as 
\be
\Delta^A_\ell=\Delta^{A\,{\rm other}}_\ell +\varepsilon_{\rm term} \Delta^{A\,{\rm term}}_\ell\,.
\ee
One can then derive the expression 
\bea
\frac{\partial C^{AB}_\ell(z_i,z_j)}{\partial \varepsilon_{\rm term}}&=& \int {\rm d}\ln k\ \bigg[ \Delta^{W_A, {\rm term}}_\ell(z_i)\ \Delta^{W_B}_\ell(z_j)  \nn\\
&&+ \Delta^{W_A}_\ell(z_i)\ \Delta^{W_B, {\rm term}}_\ell (z_j)\bigg]\ {\cal P}(k)\,.
\eea
The strategy is the same as for  the biases and $\fnl$; we compute $C_\ell(z^{\rm term}_i,z^{\rm all}_j)$ and then add its transpose,
\be
\frac{\partial C^{AB}_\ell(z_i,z_j)}{\partial \varepsilon_{\rm term}}= C^{AB}_\ell(z^{\rm term}_i,z^{\rm all}_j) + C^{AB}_\ell(z^{\rm all}_i,z^{\rm term}_j).
\ee

\label{lastpage}

\end{document}